\documentclass[letterpaper]{article} 
\usepackage{aaai25}  
\usepackage{times}  
\usepackage{helvet}  
\usepackage{courier}  
\usepackage[hyphens]{url}  
\usepackage{graphicx} 
\usepackage{natbib}  
\usepackage{caption} 
\usepackage{algorithm}
\usepackage{algorithmic}

\usepackage{newfloat}
\usepackage{listings}
\DeclareCaptionStyle{ruled}{labelfont=normalfont,labelsep=colon,strut=off} 
\floatstyle{ruled}
\newfloat{listing}{tb}{lst}{}
\floatname{listing}{Listing}
\nocopyright 

\usepackage{xspace}
\usepackage{booktabs}
\usepackage{multirow}
\usepackage{amsmath}
\def\benchname{DailyQA\xspace}

\title{DailyQA: A Benchmark to Evaluate Web Retrieval Augmented LLMs Based on Capturing Real-World Changes}
\author{
    Jiehan Cheng,
    Zhicheng Dou\thanks{Corresponding author.}
}
\affiliations{
    Gaoling School of Artificial Intelligence, Renmin University of China \\

    jiehancheng@gmail.com, dou@ruc.edu.cn
}

\usepackage{bibentry}

\begin{document}

\maketitle

\begin{abstract}

We propose DailyQA, an automatically updated dynamic dataset that updates questions weekly and contains answers to questions on any given date.
DailyQA utilizes daily updates from Wikipedia revision logs to implement a fully automated pipeline of data filtering, query generation synthesis, quality checking, answer extraction, and query classification. 
The benchmark requires large language models (LLMs) to process and answer questions involving fast-changing factual data and covering multiple domains. 
We evaluate several open-source and closed-source LLMs using different RAG pipelines with web search augmentation. 
We compare the ability of different models to process time-sensitive web information and find that rerank of web retrieval results is critical.
Our results indicate that LLMs still face significant challenges in handling frequently updated information, suggesting that DailyQA benchmarking provides valuable insights into the direction of progress for LLMs and RAG systems.
\end{abstract}

\section{Introduction}

Large language models (LLMs) has demonstrated its wide range of capabilities in the natural language processing (NLP) domain~\cite{devlin2018bert,brown2020language-gpt} and is extending its influence to more and more domains~\cite{radford2021learning,ramesh2021zero,luo2022biogpt,singhal2025toward,salinas2020deepar}. However, the world is changing fastly, and the static knowledge in the memory of LLMs is usually not updated in a timely manner~\cite{Dhingra2021TimeAwareLM}. A popular approach to this issue is to use retrieval-augmented generation (RAG)~\cite{lewis2020retrieval} techniques to provide the language model with external knowledge, allowing the model to solve problems using in-text learning methods. However, this approach often relies on external retrievers based on keyword or semantic matching~\cite{bm25,karpukhin2020dense,khattab2020colbert,chatterjee2024dreq,guo2024multimodal,mcdonald-etal-2018-deep}.
For time-sensitive queries, highly ranked documents may contain misleading information because they do not fulfill the time constraints, thus limiting the capabilities of the RAG system. So we design a time-sensitive query dataset based on realistic changes to measure the model's ability to adapt to rapidly changing information under time constraints.

\begin{figure}[t]
  \includegraphics[width=\columnwidth]{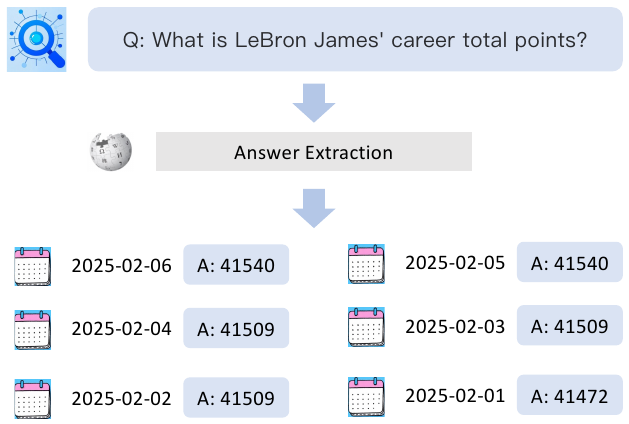}
  \caption{A example for \benchname. The answer to “LeBron James' career total points" can change every day. For each query in \benchname, we provide an answer on each day.}
  \label{fig:example}
\end{figure}

\begin{figure*}[t]
  \includegraphics[width=0.96\linewidth]{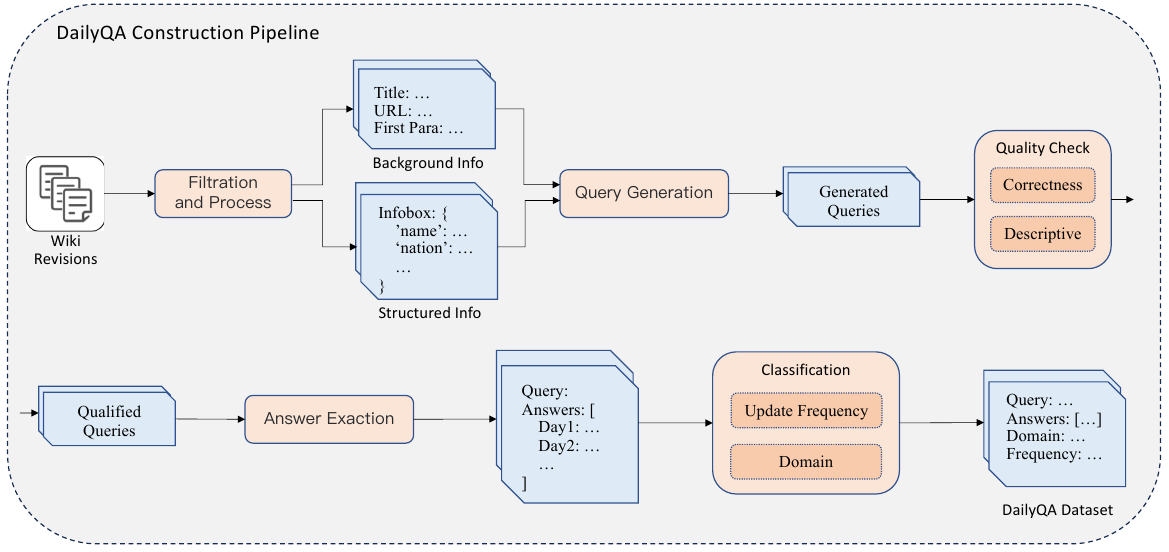}
  \caption {Overview of our \benchname dataset construction pipeline, which includes filtration and process of the raw data (Wiki revision logs), question generation, quality check, answer extraction, and query classification modules.
   In the quality check module, we check the correctness and descriptiveness of the queries. In the classification module, we classify queries based on their update frequency and domains}
  \label{fig:1}
\end{figure*}

Time-sensitive queries have been explored for a long time~\cite{10.1145/2396761.2398667,yang-etal-2024-enhancing-temporal,gade2024s,Mousavi2024timesensitive}. MRAG~\cite{MRAG} added temporal perturbations to the existing datasets TIMEQA~\cite{timeqa} and SITUATEDQA~\cite{zhang2021situatedqa} to build datasets with temporal information.  UnSeenTimeQA~\cite{Uddin2024UnSeenTimeQATQ}, in order to test the model's adherence to temporal information, constructed a virtual dataset.
However, they are both static datasets that do not reflect real-time changes in the real world, and thus do not reflect the model's ability to adapt to realistic information.
FreshLLMs~\cite{Vu2023FreshLLMsRL} manually annotated about 600 pieces of data and periodically published updated answers.
They can dynamically update the answers, but the queries are static and small in number, so the scope of real-world knowledge involved is limited. 

To investigate the capability of LLMs to adapt to complex and changing real-world knowledge, we propose a new benchmark, \benchname.
This work focuses on constructing a daily updated benchmark that contains the latest changes in reality.
As shown in Figure~\ref{fig:example}, each query in \benchname is provided with an answer on each day.
Specifically, we analyze the daily revision records of wiki pages, comparing the page versions before and after the revision and focusing on the changes to the infobox, which tends to contain concise factual information with little redundancy, and then construct the query-answer dataset using the revisions as the golden document. 
We constructed a stereo measure set by building query data that is updated weekly and corresponding answers that are updated daily. Queries can reflect changes in reality over the span of a week, and paired with the answer to that question on any given day, the time-sensitive nature of LLM can be effectively measured.
We designed a fully automated process of data filtering, query synthesis, query quality checking, and answer extraction to ensure the efficient update of the benchmark.
In experiments, we measured open-source models such as llama, Qwen, and DeepSeek-R1-Distill-Qwen based on web search augmentation. We also tested them on the DailyQA dataset under a rearrangement that takes into account both temporal and semantic, and found that the performance improved. Our contribution can be summarized as follows:
\begin{itemize}
    \item We propose \benchname, a benchmark that responds to changes in reality to measure the adaptability and time sensitivity of LLMs.
    \item We evaluated several LLMs on \benchname and proposed an improved rag method for in-context learning. Our experiments show that this task remains challenging for LLMs.
    \item We analyze further the difficulties in the task of dealing with rapidly changing real-world information, as well as the limitations to LLMs, and then propose promising research issues. 
    
\end{itemize}

\section{Related Works}

\textbf{Time sensitive QA.} There has been some work focusing on building time-sensitive benckmarks. MRAG~\cite{MRAG} builds a new benchmark on top of the existing dataset TIMEQA~\cite{timeqa} and SITUATEDQA~\cite{zhang2021situatedqa}  with temporal perturbations. TSQA~\cite{yang-etal-2024-enhancing-temporal} builds an in-domain Time sensitive dataset for nobel prize. UnSeenTimeQA~\cite{Uddin2024UnSeenTimeQATQ} builds a fictitious, contamination-free benchmark to measure the temporal reasoning ability of the model. 
However, the domains involved in these works are restricted, and the document corpus they use is static.
The scope of knowledge covered by the corpus is too small compared to the information available on the web, 
which is not a good measure of the adaptability of LLMs in the face of complex and changing information on the Internet. 
Therefore, we propose \benchname based on wiki pages covering seven domains such as science and technology, augmented with web retrieval, which is used to measure the adaptability of LLMs in the face of complex web documents.

\textbf{Realtime QA.} 
FreshLLMs~\cite{Vu2023FreshLLMsRL} manually annotates queries and publishes updated answers weekly, and they create queries of the fast change, slow change, never change, and false promising types. RealTime QA~\cite{Kasai2022RealTimeQW} also generates queries using manual annotation, and provides a platform to regularly publish queries and evaluating systems. However, the data size of the queries in the existing work is too small, which leads to a restricted domain and knowledge boundary.
For example, FreshLLMs has a fixed query set of 600 queries. RealTime QA updates about 30 queries per week.
Such amount of data is too little for reflecting changes in reality as well as for measuring and improving the performance of LLMs. We designed an automatic pipeline to update the benchmark, which can update about 3k queries data per week. And by using our script, readers can easily get answers to queries in the dataset on any given day.

\section{DailyQA Benchmark}
In this section, we introduce the DailyQA benchmark. In the following subsections, we will introduce the design principles, the build pipeline, and the data structures of DailyQA in turn. 

\subsection{Benchmark Design Principles}
DailyQA focuses on evaluating the ability of large language models to synthesize complex and changing real-world information. For this purpose, we filter and extract valuable information from daily revisions of wikipedia and use it to build a benchenmark that can be automatically updated at low cost.

To reflect the complex and changing reality, we update the set of queries in the benchmark once a week, and for each of these queries, we update its answer every day. In the evaluation phase, we give the query and specify the date, and require that the LLMs, augmented by a web search, have to correctly answer the queries of the corresponding date. 
This task is challenging and rewarding. Documents obtained through web search may be misleading because they contain information that is too old or too new, which challenges both the reranker and the LLMs. This task is valuable because in real-world scenarios, users might care about factual information about a specific day, for example, 'What is LeBron James' career score as of January 31, 2025?'

\subsection{\benchname Build Pipeline}

In this section, we describe the pipeline for building the DailyQA dataset, which includes the following parts: wiki data collection and process, query generation and quality check, and answer extraction.

\subsubsection{Wiki Data Collection and Processing}
Each time we update the query dataset, we extract all records within a week from the revision records of the wiki and filter them step by step in a rule-based approach. 
First, we only consider revisions to the main wiki page,  i.e. the page indexed by search engines, and ignore revisions to other namespaces. 
Second, we focus only on revisions in the wiki infoboxes, ignoring changes to other contents. As shown in the Appendix~\ref{sec:info_exa}, this is because wiki infoboxes tend to be well structured and purify factual information with little redundancy compared to the main text. 
Third, we process the infobox into python's dictionary format, where the content of each block in the infobox corresponds to the value of the dictionary one by one. We further filter based on key and value, that is, we remove keys of setting type such as “color1” and values of filename type such as ending with “.png”. 
For multiple changes to the same page (identified by title), we keep only the last one. We identify changes in terms of key values as the smallest unit, and for multiple changes in a single revision, we keep only one randomly to ensure the diversity of the query set and avoids multiple queries on the same entity.

After the three steps above, we filtered out infobox data that has recently been changed, has good background information (wiki body content), and is well-structured. We store the extracted value (the filtered change), the complete infobox, the title, the url, and the first paragraph of the body text in the wiki page as the extracted data units.

\begin{figure}[t]
  \includegraphics[width=\columnwidth]{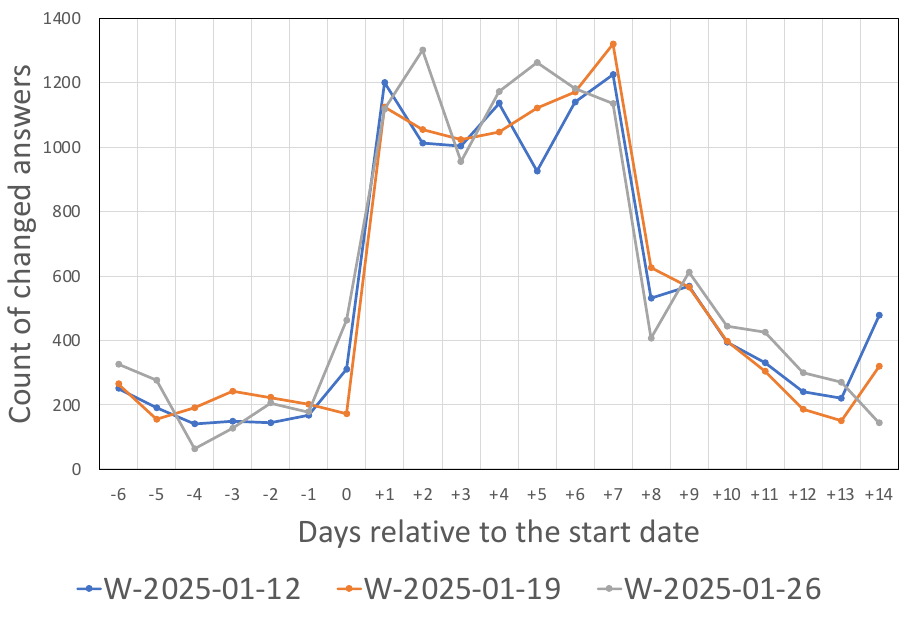}
  \caption{The number of answer changes relative to the previous day. For example, on the line with a start date of 2025/01/12, the “+1” position on the horizontal axis indicates that in the corresponding dataset, the answers for 2025/01/13 was changed by about 1,200 relative to the previous day.}
  \label{fig:answer_change_per_day}
\end{figure}

\subsubsection{Query Generation and Quality Check}

We use a LLM to automatically generate queries. We require the big model to generate a query with the extracted value in a data unit as the answer, and the infobox, the title, and the first paragraph of the body as the background information. The prompt we use is in Appendix~\ref{sec:prompt}. In this way, we fully describe the context in the query, making the query as precise as possible. At the same time, we map the answer to the value of a block of the infobox, making it easy to extract and update the answer.

In the quality checking stage, we ensure both correctness and descriptiveness of the queries. The correctness of a query means that the query should be able to be answered accurately when sufficient information is provided. 
The descriptiveness of a query means that the query should be reasonably worded and clearly present the background of the problem, without relying on the background information provided to the LLM. 

To check the correctness, we provide the original wiki title, the first paragraph of the body, and the infobox in the data unit as references, and ask the LLM to answer the question based on them. We treat the query as a valid one if the sub match metric between the model answer and the ground truth is 1. 
To check the descriptiveness, we use DuckDuckGo search api to get the top 10 results and keep only the queries that can successfully retrieve the corresponding wiki pages. 
In this way, we use external knowledge anchoring to avoid semantic bias.

After the above process, through automatic query generation and quality checking, we automatically obtain a set of correct and descriptive queries that reflect the changes of the reality.
On average, we filtered out about 8,000 valid queries from about 11,000 queries per update.

\subsubsection{Answer Extraction}
According to the above processes, the answer to the question is set to the value of a certain block in the infobox. Therefore, we only need to monitor the corresponding page, the corresponding infobox and the corresponding block every day to get its value to get the answer updated every day. Specifically, we can find the revision history of a page from the wiki logs, and get the correct answer based on the last revision before the requested date. This approach makes it possible to get the answer to a query in the dataset on any day at a very low cost.

\begin{figure}[t]
  \includegraphics[width=0.96\columnwidth]{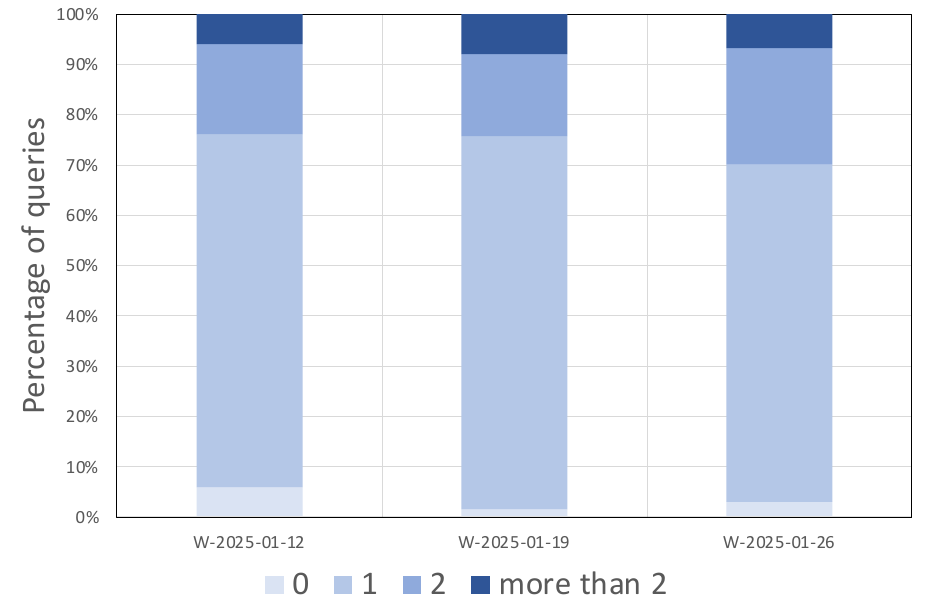}
  \caption{Percentage of queries with different answer change times. For example, as shown in the left bar, in the query dataset for 2025/01/12-2025/01/18, the percentage of queries whose answer change once is about 70\%. Note that consistent with Figure~\ref{fig:answer_change_per_day}, we count answer changes over a three-week period that includes the week before and after.}
  \label{fig:answer_change_per_query}
\end{figure}

\subsubsection{Classification}
In this section, we introduce the classification of query types for QA datasets. We classify the queries in two perspectives, including their update frequency and domain.

\textbf{Update frequency.} We use the update frequency to mark how often the answer to a query changes. 
As shown in Figure~\ref{fig:answer_change_per_day}, we statistic the day-by-day variation of answers in the dataset for three updates. 
Each line in the graph represents an weekly update of queries, for example, “W-2025-01-12” means that this update corresponds to the week starting from 2025-01-12.
In the figure, we take the first day of the corresponding week as the start date (day 0 on the horizontal axis). We use the horizontal axis to indicate the nth day relative to the start date and use the vertical axis to represent the number of changed answers on that day relative to the previous day. 
We observe the changes in the answers over a three-week time span and find that the answers change a great deal on day 1-7, while significantly smaller in the week before and after. 
Since the query is based on the variation on day 1-7, this is as expected. 
We label queries that do not change from day 8 as "infrequent\_update" and the others as "frequent\_update".

Above we present the distribution of queries in the dataset at a macro level, and below we will present query-by-query statistics to make it easier for readers to filter and use the parts of their interest. 
We count the number of answer changes for each query over a three-week period and present it in Figure 3. 
It can be seen that the number of changes of answers for most of the questions is in the range of 0-1 times, and there are also frequent changes of answers. 
When answers change infrequently, the difficulty of the queries decreases significantly because web documents tend to include less misleading information. Readers can filter the dataset and use the parts of interest according to their desired difficulty of the task.

\begin{figure}[t]
  \includegraphics[width=0.96\columnwidth]{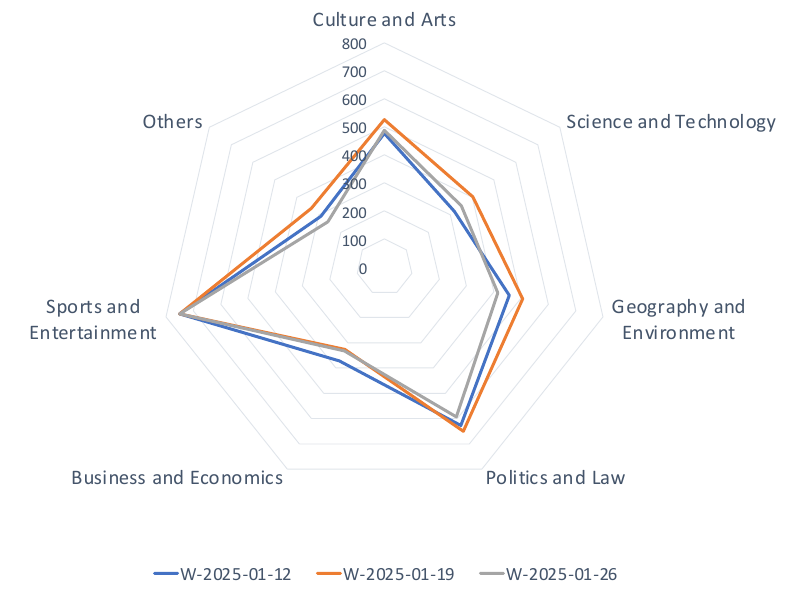}
  \caption{Distribution of the queries in different domains. In the labels, "W-2025-01-12", for example, means a query update corresponds to the week starting from 2025-01-12.}
  \label{fig:leida}
\end{figure}

\textbf{Domains.} 
Our query set is based on the comprehensive Wikipedia, so it covers multiple fields. We classify the query set by domain so as to provide convenience for in-domain researches. 
We use a LLM to classify queries into 7 classes by means of zero-shot, including Science and Technology, Culture and Arts, Geography and Environment, Politics and Law, Business and Economics, Sports and Entertainment, Others.
We categorize queries that do not belong to the first six domains as Others.
In order to balance the distribution of queries across different domains, we set the maximum number of queries in each domain in each update to 750 and  use non-repetitive random sampling method to shrink the oversized query set.
The distribution of the data in the different classes is shown in Figure~\ref{fig:leida}. Readers are free to choose the domains of their interest.

\subsection{\benchname Data Structure}

After the above pipeline, the data structure of our DailyQA is as follows: 
\begin{itemize}
    \item DailyQA adds a new query dataset every week, which is based on factual information about the latest changes in reality. Our Pipeline automatically crawls the data, generates the query, and check the quality.
    \item Each query is paired with its update frequency, domain, and golden document (i.e., a Wiki page), which consists of the title, the url, the first paragraph of the body, and a dictionary-formatted infobox.
    \item Each query's answer is updated daily, which means it has a corresponding answer on any given date. In fact, we provide a script for extracting answers that helps users to obtain answers for a given date easily and cheaply.
\end{itemize}

\begin{table*}[t]
  \centering
  \small
  \begin{tabular}{llcccccccc}
    \toprule
    & & \multicolumn{4}{c}{\textbf{Search w/ Time}} & \multicolumn{4}{c}{\textbf{Seach w/o Time}} \\
    \cmidrule(lr){3-6}
    \cmidrule(lr){7-10}
    \textbf{LLM} & \textbf{Pipeline} &  \textbf{SM} & \textbf{Rouge-L} & \textbf{F1} & \textbf{Acc} & \textbf{SM} & \textbf{Rouge-L} & \textbf{F1} & \textbf{Acc} \\
    \midrule
    \multirow{5}{*}{Qwen2.5-72B-Instruct} & w/o Search & 0.120 & 0.011 & 0.021 & 0.139 & 0.120 & 0.011 & 0.021 & 0.139\\
    & Snippet & 0.242 & 0.159 & 0.185 & 0.249 	& 0.263	& 0.200 & 0.226 & 0.286\\
    & Doc & 0.356 & 0.241 & 0.275 & 0.364 	& 0.479	& \underline{0.373} & \underline{0.410} & 0.492\\
    & Rerank  & \textbf{0.392} & \textbf{0.308} & \textbf{0.338} & \textbf{0.416} & \textbf{0.502} & \textbf{0.413} & \textbf{0.446} & \textbf{0.513}\\
    & Rerank-T  & 0.311 & \underline{0.242} & \underline{0.268}& 0.324 & 0.311 & 0.250 & 0.276 & 0.366\\
    \midrule
    Qwen2.5-7B-Instruct  & \multirow{2}{*}{Rerank} & 0.350 & 0.165 & 0.205& 0.366 & 0.444 & 0.216 & 0.264 & 0.457\\
    Qwen2.5-32B-Instruct & & 0.351 & 0.194 & 0.242 & 0.364 & 0.447 & 0.255 & 0.312 & 0.455 \\
    \midrule
    DSRD-Qwen-32B & \multirow{2}{*}{Rerank} & 0.363 & 0.101 & 0.155 & 0.379 & 0.433 & 0.122 & 0.181 & 0.452 \\
    GPT-4o-mini &  & \underline{0.381} & 0.209 & 0.252 & \underline{0.403} & \underline{0.484} & 0.268 & 0.317 & \underline{0.498} \\
    \bottomrule
  \end{tabular}
  \caption{Evaluation on \benchname with different retrieval methods, RAG pipelines, and LLMs. \textit{Search w/ Time} means web searching queries with dates, and \textit{Search w/o Time} means web searching raw queries. In the RAG pipeline, \textit{w/o Search} means no Web Retrieval Augmentation, \textit{Snippet} means using the web-retrieved snippets as reference, \textit{Doc} means using documents crawled via URLs, \textit{Rerank} means reranking the documents, \textit{Rerank-T} means reranking the documents based on relevance and time.   The best results are in \textbf{bold} and the second-best are \underline{underlined}.}
  \label{tab:main}
\end{table*}

\begin{table}[!t]
    \centering
    \small
    \begin{tabular}{lcc}
    \toprule
    \textbf{Model} & \textbf{SM1} & \textbf{SM2} \\
    \midrule
    Qwen2.5-72B-Instruct  & \textbf{0.302} & \textbf{0.693} \\
    Qwen2.5-32B-Instruct  &0.269 & 0.681 \\
    DSRD-Qwen-32B & 0.253 & 0.677\\
    GPT-4o-mini & 0.291 & 0.688 \\
    \bottomrule
    \end{tabular}
    \caption{SM of the LLMs on frequent\_update (SM1) and infrequent\_update (SM2) queries. We use the pipeline of \textit{Rerank} and \textit{Search w/o Time} for all the LLMs.}
    \label{tab:freq}
\end{table}

\begin{table}[!t]
    \centering
    \small
    \begin{tabular}{lcc}
    \toprule
    \textbf{Model} & \textbf{SM} & \textbf{Acc} \\
    \midrule
    Qwen2.5-72B-Instruct  & 0.466 & 0.482 \\
    Qwen2.5-32B-Instruct  &0.445 & 0.458 \\
    DSRD-Qwen-32B & 0.434 & 0.449\\
    GPT-4o-mini & \textbf{0.477} & \textbf{0.489} \\
    perplexity.ai & 0.471 &  0.485\\
    \bottomrule
    \end{tabular}
    \caption{Performance of different LLMs on the DailiyQA dataset in the Science and Technology domain. Except perplexity.ai, we use the pipeline of \textit{Rerank} and \textit{Search w/o Time} for all the LLMs. For perplexity.ai, we provide queries with the specified date and require the service to search for and answer the queries autonomously.}
    \label{tab:closed_source}
\end{table}

\begin{table*}[t]
  \centering
  \small
  \begin{tabular}{lccccccc}
    \toprule
    \textbf{Model} &  \textbf{ST}  & \textbf{CA} & \textbf{GE} & \textbf{PL} & \textbf{BE} & \textbf{SE} & \textbf{Ot} \\
    \midrule
    Qwen2.5-72B-Instruct & 0.466 & \textbf{0.541} & \textbf{0.560} & \textbf{0.580} & \textbf{0.421} & \textbf{0.442} & \textbf{0.474} \\
    Qwen2.5-32B-Instruct & 0.445 & 0.517 & 0.532 & 0.561 	& 0.372	& 0.413 & 0.425 \\
    DSRD-Qwen-32B & 0.434 & 0.470 & 0.501 & 0.497 	& 0.315	& 0.408 &  0.348\\
    GPT-4o-mini& \textbf{0.477} & 0.512 & 0.538 & 0.548 	& 0.394	& \textbf{0.442} & 0.447\\
    \bottomrule
  \end{tabular}
  \caption{SM of the LLMs on \benchname in seven domains, including Science and Technology (ST), Culture and Arts (CA), Geography and Environment (GE), Politics and Law (PL), Business and Economics (BE), Sports and Entertainment (SE), Others (Ot).}
  \label{tab:domains}
\end{table*}
\section{Experiments}
\subsection{Baselines}

We measured the performance of the RAG system on \benchname with different web retrieval methods, rag pipelines, and LLMs. 

We use \textbf{Search w/ Time} and \textbf{Search w/o Time} to denote different web search methods. 
The former means that we add the required date to the query and retrieve the query with the date it over the web, while the latter means that we retrieve the query without the date over the web. By comparing these two methods, we found out the limitations of solving \benchname directly with the help of search engines.

We compare several types of RAG pipelines. As shown in Table~\ref{tab:main}, \textbf{w/o Search} means that we do not rely on any information retrieved from the web and only rely on the LLM to answer the questions. \textbf{Snippet} means that we use the web snippet retrieved by the search engine as the reference, and provide it to the LLM in the order of the web search to help answer the questions. 
\textbf{Doc} means that we obtain the html page based on the URLs returned from the web search, and extract the text of the pages. We then provide them to LLMs in the order of the web search to help answer the questions.
\textbf{Rerank} denotes that based on the documents of the html pages, we chunk and rerank them, and then provide them to LLMs in the order of reranking. 
\textbf{Rerank-T} means reranking documents based on relevance and time. Based on the "Rerank" pipeline above, we further rerank the chunks with the update time.
Specifically, based on the topk document chunks from "rerank", we prioritize the documents whose modification date is before the query date and closer to the query date. 
By this heuristic approach, we try to provide assistance to LLM in identifying the correct reference documents in the rerank phase.

As shown in Table~\ref{tab:main}, we evaluated different kinds and sizes of LLMs. Qwen-2.5~\cite{yang2024qwen2} series is a set of powerful large language models developed by Qwen that showcase advanced capabilities in natural language understanding and generation. We use Qwen-2.5-72B-Instruct as the base model to evaluate the performance of different rag pipelines. We use the Qwen-2.5 series of models to evaluate the impact of model scale. 
For closed-source models, we measured the performance of GPT-4o-mini~\cite{achiam2023gpt}.
Deepseek-r1~\cite{guo2025deepseek} is the latest and one of the state-of-the-art LLMs for universal large models.
For cost reasons, we measured the performance DeepSeek-R1-Distill-Qwen-32B instead of Deepseek-r1.
We use “DSRD-Qwen-32B” to represent the DeepSeek-R1-Distill-Qwen-32B model.

\subsection{Metrics}
We use the rule metrics and the model evaluation metrics. For the rule metrics, we use subset match (SM), Rouge-L, and F1. The value of SM is 1 if the correct answer is in the prediction and 0 otherwise. For the model evaluation metrics, we require GPT-4o to determine the accuracy (Acc) of the predicted answers. 
Specifically, we asked the LLM to score the similarity of the predicted results to the standard answers, with 5 being completely similar and 1 being completely irrelevant. We computed four and five as correct, i.e., Acc of one, and computed the others as Acc of zero.

\subsection{Implementation Details}

In the dataset construction phase, we use the pywikibot packet to download and process Wiki logs, and we use Qwen-72B-Instruct to generate queries. 
We use Qwen-72B-Instruct to answer the queries with golden references for quality check. 
In the evaluation phase, we use the api of DuckDuckGo as the web search engine, and  use Trafilatura to extract the main text in the HTML. 
We manually specified seven domains and used Qwen-72B-Instruct to identify the domain to which the queries belong. 
In retrieval enhancement, we uniformly use top 12 snippets, documents or chunks as the reference and use bge-v2-m3 as the reranker.
We evaluate on the query update corresponds to the week starting
from 2025-01-12, specify the query date as 2025-01-19. We use GPT-4o for the model evaluation. 
\section{Results}

\subsection{Main Results}

For Qwen2.5-72B-Instruct, \textbf{web retrieval is necessary on \benchname and reranking the raw web-retrieved documents can effectively improve performance}. As shown in Table~\ref{tab:main}, the results show that the model without web search performs substantially worse than others. This is consistent with our expectations since \benchname is constructed based on fresh information. 
Using the original web text is more helpful than using snippets from the search engine, and reranking the raw web-retrieved documents instead of the web retrieval order further improves performance. This suggests that in order to solve this task, we need to keep digging deeper and pay attention to the details of the retrieved content,  rather than relying only on summaries. This challenges the information integration capabilities of LLMs and the design of RAG pipelines.

\textbf{Increasing the scale of the model helps a lot in the metrics of Rouge-L and F1 on \benchname}. The results for different sizes of Qwen2.5 models in the Table~\ref{tab:main} show that increasing the model scale leads to a weak improvement in SM and ACC, and a significant improvement in Rouge-L and F1. 
This means that as the model scale increases, the model tends to be able to answer questions in shorter words, which reflects that the model's answer is more concise with less redundancy.
Increasing the scale of LLMs enhances the ability to process time-sensitive realistic documents. 
It confirms the challenges of \benchname for LLMs and also illustrates the ability of LLMs to find the required details in complex web references.

\textbf{Qwen2.5-72B-Instruct works best on \benchname on all the metrics}. We compared several open-source and closed-source models and found that Qwen2.5-72B-Instruct performs best. It outperforms over models on all the metrics. 
Notably, Qwen2.5-32B-Instruct outperforms DSRD-Qwen-32B on most metrics. DSRD-Qwen-32B, which has been validated to have stronger inference, does not perform as well as the same-sized Qwen2.5-32B-Instruct on this benchmark. This shows that its capability to  extract document details is degraded, as well as the possibility of more serious hallucinations. It's suggested that our benchmark is complementary to the LLM evaluation, in the dimension different from the reasoning ability, thus helping to measure LLM's ability more comprehensively.

\textbf{Our preliminary attempts to integrate time information in the RAG pipeline does not result in a performance improvement.} Specifying the date in the web retrieval module and adding time information to the rerank both have a negative effect on the performance.
As shown in Table~\ref{tab:main}, the performance of Search w/o Time is weakly bertter than that of Search w/ Time, and the performance of Rerank is better than that of Rerank-T. This shows that Adding time descriptions directly to the query or rerank the chunks based on time did not result in an improvement.
The reason may be that the search engine is not able to accurately understand the intent and process the complex queries so as to return the correct document.
This suggests a challenge in calling the search engine more accurately when dealing with time-sensitive real-world problems. 
Precise retrieval through agentic RAG may be a promising approach in the future.

\textbf{All models perform better on infrequent update queries than on frequent update queries}. As shown in Table~\ref{tab:freq}, We analyze the accuracy of the model on problems with different frequencies of change. The results show thar all models have lower accuracy on the frequent update queries. They are more difficult because documents retrieved from the website tend to include more misleading information, which challenges the ability of LLMs to reason and make temporal judgments. 
Notice that the gap in model performance is larger on frequent updated quries than on infrequent update quries. This suggests that frequently updated queries are more difficult and that there is more potential for the model to improve on such queries.

\textbf{\benchname is a challenge for existing web retrieval augmented LLM services}. 
To measure the  difficulty of queries in \benchname, we measured the it on perplexity.ai and compared it with our methods.
As shown in Table~\ref{tab:closed_source}, perplexity.ai's accuracy on the dataset is comparable to that of our rerank method and there are still about half of the queries that the model cannot answer correctly. 
This shows that \benchname benchmark is still a challenge for existing industry solutions.

\subsection{Results in Multiple Domains}

In order to introduce \benchname in more detail and to judge the difficulty of the queries in different domains, we measure the accuracy of the models in different domains.
The difficulty of the questions varies from one area to another. All the models are relatively more accurate in the domains of Culture and
Arts (CA), Geography and Environment (GE), Politics and Law (PL), while they are relatively less accurate within other domains. This is because content in these fields tends to be updated infrequently, while in other fields such as
Sports and Entertainment (SE), questions like “What is LeBron James' career total points?” tend to be updated frequently, thus posing a greater challenge.

We find that different series of models have their own areas of specialization. Although Qwen2.5-72B-Instruct has the best overall performance, it does not achieve the best results in all domains. Gpt-4o-mini performs better than Qwen2.5-72B-Instruct in Science and Technology (ST) domain. This implies that due to the different training data and methods, the LLMs may have their own good and bad areas. This provides motivation for building multi-model collaborative agents to solve cross-domain problems.

\subsection{Analyse of Challenges}
By measuring the performance of the open-source and closed-source LLMs on our benchmark, we can evaluate the ability of these LLMs to process time-sensitive web information. The challenge of this task is mainly twofold. 

First, web information is complex and diverse, and it is worth exploring how to fully utilize search engines to obtain the needed information. As shown in Table~\ref{tab:main}, the modification of adding timestamps by rules may not achieve the expected results, so invoking search engines by issuing queries with the help of LLMs may be a promising direction. 

Second, the information in the related documents is time-sensitive. Although the reranked documents have similar semantics with queries, they are likely to contain information that does not meet the time requirement and cause misleading. We have explored methods to rank web pages based on their modification date but it did not result in improvements, possibly because the modification time of a web page is not equivalent to the effective time of the information, and many web pages lack the information of the modification time.
Therefore, comprehensively analyzing the retrieved documents and obtaining time information based on content may be a promising direction.

\section{Conclusions}
We propose DailyQA, a benchmark reflecting changes in reality, to measure LLMs' adaptability and time sensitivity to factual information. We perform the experiments using both open-source and closed-source models and the results show that this task remains a challenge for existing solutions. We further analyze the difficulties in the task of dealing with rapidly changing real-world information, as well as the limitations to LLMs. 
We expect that by solving the queries in \benchname, the capabilities of LLMs can be further refined and released.

\section*{Limitations}
Our benchmark is intended to evaluate the LLMs' ability to process Internet information, and does not focus on the LLMs' logical reasoning ability. Therefore, our dataset contains only one-hop queries and does not include multi-hop queries or false-premising queries.

Due to the limited resources, we did not evaluate the state-of-the-art LLMs such as GPT-o1, DeepSeek-R1, etc. We leave the evaluations on these models for future work.

Affected by the diversity of web page structures, in our implementation, we failed to get the information of the update time for a portion of the web page , so this may degrade the performance of our \textit{Rerank-T} pipeline.

\section*{Ethics}

This paper constructs a benchmark dataset derived from Wikipedia pages and LLM-generated queries. While Wikipedia content may reflect the personal biases of its contributors, and recently updated pages could occasionally contain unverified information, our methodology mitigates these limitations by exclusively utilizing structured infoboxes as the data source. This approach significantly reduces subjective statements in the referenced Wikipedia texts. Similarly, while LLM outputs may inherit potential biases from training data, our quality check process for queries serves as an inherent quality control mechanism. The dataset is expressly designed for academic benchmarking in NLP research, not for commercial applications. All Wikipedia-derived content remains subject to its original CC BY-SA license terms.

\clearpage
\bibliography{Cheng}

\clearpage

\appendix

\section{An Infobox Example}
\label{sec:info_exa}

\begin{figure}[h]
  \includegraphics[width=\columnwidth]{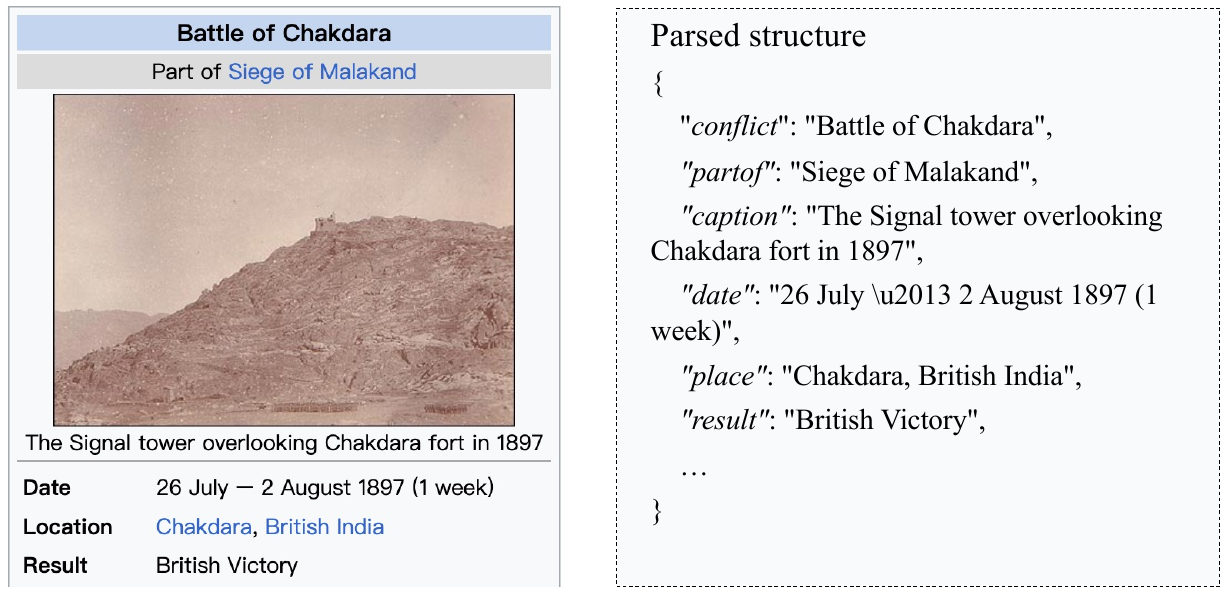}
  \caption{An example of an infobox from a wikipedia page (left), and its processed data structure (right). The infobox is from the wikipedi \textit{https://en.wikipedia.org/wiki/Battle\_of\_Chakdara.}}
  \label{fig:infobox_exa}
\end{figure}

As shown in Figure~\ref{fig:infobox_exa}, we introduce An example of an infobox from a wikipedia page, and its processed data structure.
We focus only on the infobox structure in the wikipedia page in data processing, and process it into a python dictionarywith the help of the pywikibot tool, which facilitates the information extraction and the understanding of LLMs in the query generation process.

\section{Prompts}
\label{sec:prompt}
\subsection{Prompt for Query Generation}
\begin{figure}[h]
  \includegraphics[width=\columnwidth]{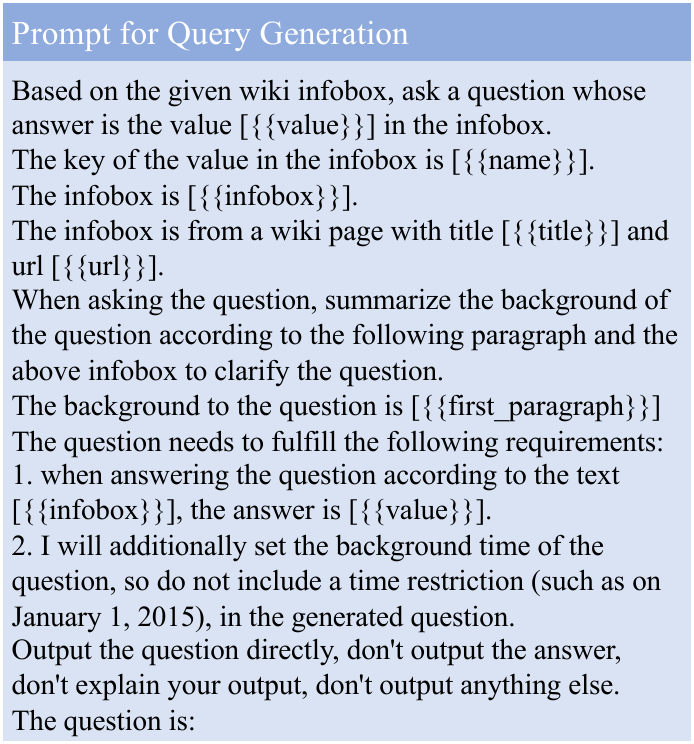}
  \caption{Prompt for Query Generation}
  \label{fig:p_querygen}
\end{figure}

\newpage
\subsection{Prompt for RAG}
\begin{figure}[h]
  \includegraphics[width=\columnwidth]{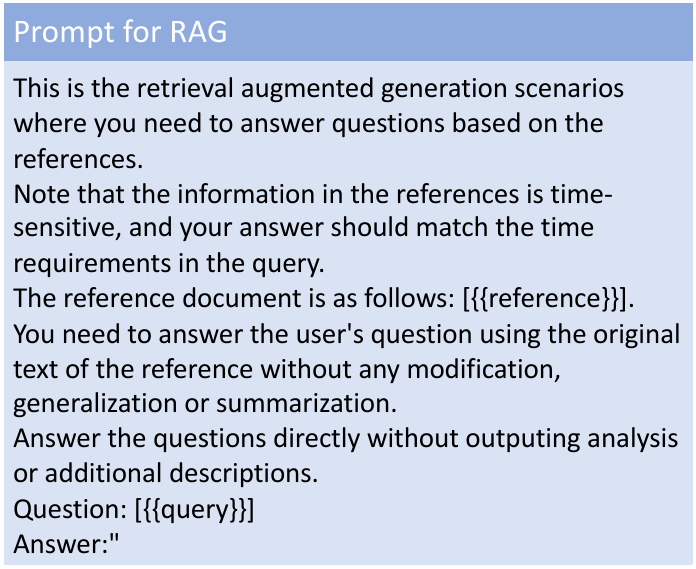}
  \caption{Prompt for RAG}
  \label{fig:p_rag}
\end{figure}

\section{Examples for Generated Queries}

\begin{figure}[h]
  \includegraphics[width=\columnwidth]{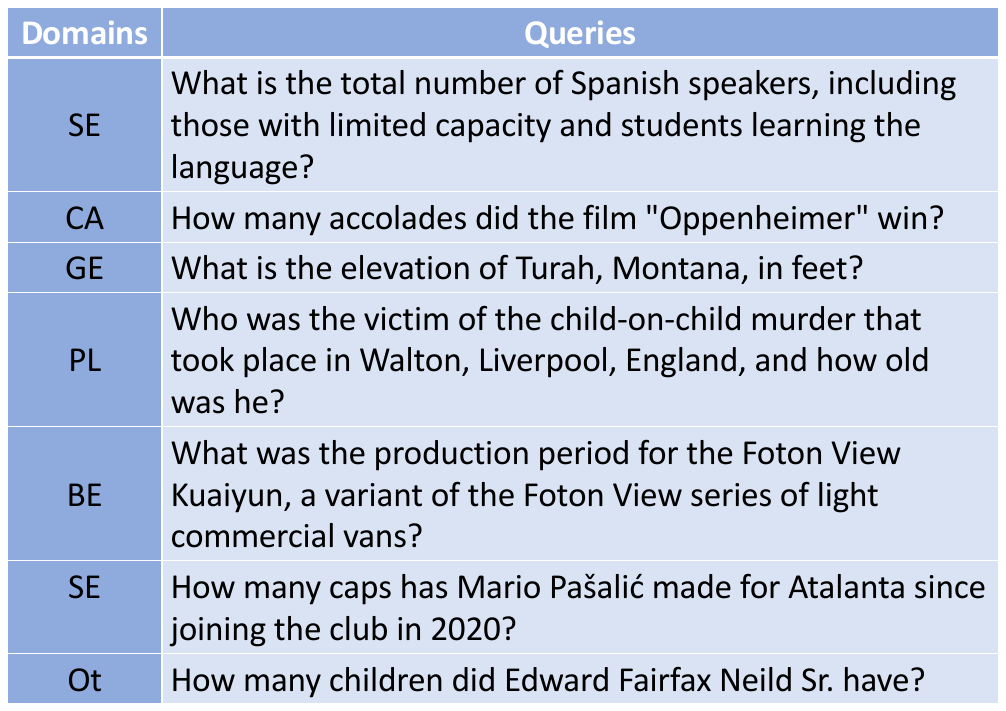}
  \caption{Examples for generated queries in different domains, including Science and Technology (ST), Culture and Arts (CA), Geography and Environment (GE), Politics and Law (PL), Business and Economics (BE), Sports and Entertainment (SE), Others (Ot).}
  \label{fig:queries}
\end{figure}

\end{document}